# Quantum anomalous Hall effect from intertwined moiré bands

Tingxin Li[1,2], Shengwei Jiang[2,3], Bowen Shen[1], Yang Zhang[4], Lizhong Li[1], Trithep Devakul[4], Kenji Watanabe[5], Takashi Taniguchi[5], Liang Fu[4], Jie Shan[1,3,6*], Kin Fai Mak[1,3,6*]

[1]School of Applied and Engineering Physics, Cornell University, Ithaca, NY, USA
[2]Key Laboratory of Artificial Structures and Quantum Control (Ministry of Education), School of Physics and Astronomy, Shanghai Jiao Tong University, Shanghai, China
[3]Laboratory of Atomic and Solid State Physics, Cornell University, Ithaca, NY, USA
[4]Department of Physics, Massachusetts Institute of Technology, Cambridge, MA, USA
[5]National Institute for Materials Science, 1-1 Namiki, 305-0044 Tsukuba, Japan
[6]Kavli Institute at Cornell for Nanoscale Science, Ithaca, NY, USA

Email: jie.shan@cornell.edu; kinfai.mak@cornell.edu
These authors contributed equally: Tingxin Li, Shengwei Jiang, Bowen Shen, Yang Zhang.

**Electron correlation and topology are two central threads of modern condensed matter physics. Semiconductor moiré materials provide a highly tunable platform for studies of electron correlation [1-12]. Correlation-driven phenomena, including the Mott insulator [2-5], generalized Wigner crystals [2, 6, 9], stripe phases [10] and continuous Mott transition [11, 12], have been demonstrated. However, nontrivial band topology has remained elusive. Here we report the observation of a quantum anomalous Hall (QAH) effect in AB-stacked MoTe$_2$/WSe$_2$ moiré heterobilayers. Unlike in the AA-stacked structures [11], an out-of-plane electric field controls not only the bandwidth but also the band topology by intertwining moiré bands centered at different high-symmetry stacking sites. At half band filling, corresponding to one particle per moiré unit cell, we observe quantized Hall resistance, $h/e^2$ (with $h$ and $e$ denoting the Planck's constant and electron charge, respectively), and vanishing longitudinal resistance at zero magnetic field. The electric-field-induced topological phase transition from a Mott insulator to a QAH insulator precedes an insulator-to-metal transition; contrary to most known topological phase transitions [13], it is not accompanied by a bulk charge gap closure. Our study paves the path for discovery of a wealth of emergent phenomena arising from the combined influence of strong correlation and topology in semiconductor moiré materials.**

Two-dimensional moiré heterostructures of van der Waals materials present a new paradigm for engineering electron correlation, topology, and their interplay [8, 14-16]. In graphene systems, moiré patterns can produce topologically nontrivial bands with valley-contrasting Chern numbers to enforce time-reversal symmetry of the single-particle band structure [14]. With sufficiently flat bands, correlation-driven states with broken symmetry are favored. Orbital ferromagnetism and the QAH effect have been reported [17-19] following the initial discovery of superconductivity and correlated insulating states in graphene moiré systems [20]. Conversely, in semiconducting transition metal dichalcogenide (TMD) heterobilayers, the moiré bands are topologically trivial [1, 7]. The low-energy physics is dominated by strong electron correlation; it is well described by a single-band Hubbard model [1, 7], endowed by broken inversion symmetry and Ising spin anisotropy in monolayer TMDs. Intertwining two copies of electronic states in twisted TMD homobilayers has been envisioned to realize a generalized Kane-Mele model for interacting quantum particles [21, 22]; an abundance of exotic states of matter, including the topological Mott insulators, QAH insulators and chiral spin liquids, has been predicted [23-26]. To date, nontrivial band topology has not been realized in TMDs.

In this study, we report an electric-field-induced topological phase transition from a Mott insulator to a QAH insulator in near-60-degree-twisted (or AB-stacked) MoTe$_2$/WSe$_2$ heterobilayers. The two TMDs form a triangular moiré superlattice with three high-symmetry stacking sites: MM, MX and XX (M = Mo, W; X = Te, Se) (Fig. 1a). The lattice mismatch between the two materials is about 7%, giving rise to a moiré lattice period of $\sim$ 5 nm and a moiré density of $n_M \approx 5 \times 10^{12}$ cm$^{-2}$ (Ref. [11]). In each monolayer, the band edges are located at the K/K' point of the Brillouin zone with two-fold (coupled) spin-valley degeneracy [1, 21]. The heterobilayer has a type-I band alignment with both conduction and valence band edges from MoTe$_2$, and a valence band offset of 200-300 meV [11]. The application of an out-of-plane electric field can reduce the band offset and tune the moiré band structure because interlayer tunneling contributes substantially to the formation of moiré potential. We have recently reported a continuous bandwidth-tuned Mott transition in near-0-degree-twisted (or AA-stacked) MoTe$_2$/WSe$_2$ at fixed filling of one hole per moiré unit cell ($\nu = 1$) (Ref. [11]). We did not observe any topological phase transitions there. Important distinctions between the two stacking orders include different high-symmetry stacking sites and interlayer spin alignment; the latter strongly influences interlayer tunneling and the moiré band structure (Methods).

We perform magneto-transport measurements on dual-gated Hall bar devices of AB-stacked MoTe$_2$/WSe$_2$ heterobilayers down to 300 mK (Methods). Figure 1c and 1d show, respectively, the longitudinal resistance ($R_{xx}$) and the magnitude of the Hall resistance ($R_{xy}$) of device 1 as a function of top gate voltage ($V_{tg}$) and bottom gate voltage ($V_{bg}$) at 300 mK and at zero magnetic field. The two gate voltages independently tune the filling factor ($\nu$) and the vertical electric field ($E$), averaged between regions above and below the heterobilayer. At small $E$, we observe two prominent resistance peaks at $\nu = 1$ and $\nu = 2$, corresponding to the Mott and the single-particle moiré band insulating states, respectively. As $E$ increases, $R_{xx}$ of both states drops by orders of magnitude. These behaviors are similar to that observed in AA-stacked devices [11]. However, there is a peculiar region (which is absent in AA-stacked devices) near the boundary of the $\nu = 1$ insulating state with $R_{xx}$ substantially lower than in the nearby metallic state (marked by dashed circle). Concomitantly, the region exhibits nonzero Hall resistance close to 26 k$\Omega$ (Fig. 1d). This signals the arrival of a new state with broken time-reversal symmetry.

We characterize the state in depth by parking the gate voltages at the region of minimum $R_{xx}$ at zero magnetic field. Figure 2a and 2b show $R_{xy}$ and $R_{xx}$ as a function of out-of-plane magnetic field at temperature ranging from 300 mK to 8 K. At low temperature, we observe sharp magnetic hysteresis with coercive field around 10 mT; $R_{xy}$ is nearly quantized at $h/e^2$ and $R_{xx} \approx 0.05 h/e^2$ is small at zero magnetic field ($\frac{h}{e^2} \approx 25.8$ kΩ is the resistance quantum). As temperature increases, we observe both a suppression of hysteresis and a departure from resistance quantization in $R_{xy}$. This is accompanied by a rapid increase in $R_{xx}$, which also exhibits negative magnetoresistance. Hysteresis vanishes by 8 K; $R_{xy}$ depends linearly on magnetic field and $R_{xx}$ becomes nearly field-independent. We summarize the temperature dependence of the zero-field magnitude of $R_{xy}$ and $R_{xx}$ in Fig. 2c to better identify the transitions. Particularly, the Hall resistance remains quantized up to about 2.5 K, and stays finite up to 5-6 K, the Curie temperature for magnetic ordering.

These observations are distinctive experimental signatures of a quantized anomalous Hall effect [27-30]. The slight imperfection (in both $R_{xy}$ quantization and residual $R_{xx}$) arises presumably from remnant dissipation in the bulk and imperfect electrical contacts, which remain a technical challenge for semiconductor moiré materials. The transition temperatures in AB-stacked MoTe₂/WSe₂ for both resistance quantization and magnetic ordering are comparable to that in twisted bilayer graphene [17]. The QAH effect here is robust and reproducible. We have studied a total of five devices. All show similar behaviors: three exhibit the QAH effect at $\nu = 1$ (Extended Data Fig. 2); the rest devices show quantized Hall resistance under a moderate magnetic field of about 1 T, likely due to higher disorder levels.

The new state is a QAH insulator, in which electrical transport is dominated by chiral edge states [30]. The insulator possesses nontrivial band topology with a total Chern number of occupied bands, $c = 1$, inferred from the quantized Hall resistance ($R_{xy} = \frac{1}{c}\frac{h}{e^2}$). This is further verified by studying the dispersion of the state in doping density and magnetic field and invoking the Streda formula, which relates the Hall resistance to the derivative of doping density with respect to magnetic field [31]. Figure 2d shows $R_{xx}$ as a function of density $\nu$ (in the unit of $n_M$) and magnetic field $B$. We choose a slightly higher temperature (4.1 K) to suppress the Landau levels in MoTe₂ under high magnetic field (see Extended Data Fig. 3 for result at 300 mK); the QAH state is not fully developed at low field ($< 1$ T) at this temperature. We identify the QAH state by the resistance minimum (dashed line) and determine $c = \frac{h}{e}\left|\frac{d\nu}{dB}\right| n_M = 0.95 \pm 0.05$ from the slope of the dashed line. The result is fully consistent with the quantized Hall resistance.

Next we examine the transition from a topologically trivial Mott insulator to a topologically nontrivial QAH insulator under an electric field at $\nu = 1$. Figure 3a shows the electric-field dependence of $R_{xx}$ under zero magnetic field for varying temperatures. No hysteresis is observed down to 300 mK (Extended Data Fig. 4). We identify three distinctive regions for $E$. The temperature dependence of $R_{xx}$ for one representative electric field from each region is shown in Fig. 3b. For small $E$ (shaded in orange), we observe large $R_{xx}$, which diverges rapidly as temperature decreases. This is a Mott insulator state, in agreement with a previous study [11]. At large $E$ (shaded in pink), resistance drops to about 10 kΩ at the highest $E$ available in this study; it has weak temperature dependence and is finite in the zero-temperature limit. This corresponds to a metallic state,. In the middle region (shaded in blue), $R_{xx}$ exhibits an insulating behavior above ~ 5 K and drops rapidly below it, as the QAH insulator is stabilized and transport is dominated by the chiral edge states. These results show that the Mott-to-QAH insulator transition precedes the insulator-to-metal transition.

We extract the charge gap in the insulating states (circles and top axis in Fig. 3c) from an activation fit of the temperature dependence of the resistance (Extended Data Fig. 5). Inside the QAH phase the activation fit can only be performed in a narrow temperature range. To directly access the bulk charge gap across the topological phase transition without influence from the chiral edge states, we also perform electronic compressibility measurements [32] (Methods). Figure 3d shows the gate voltage dependence of the differential capacitance, $C$, between the top gate electrode and the sample of device 6, at 300 mK and at zero magnetic field. The capacitance is normalized by the geometrical capacitance, $C_g$, defined by the sample-gate distance and the dielectric constant of the gate dielectrics. It probes the electronic compressibility of the sample: $C/C_g \approx 0$ when the sample is charge neutral (not shown); $C/C_g \approx 1$ when the sample is heavily hole-doped and behaves like a metallic plate [33]. Each dip in $C/C_g$ at finite filling factors corresponds to an incompressible/insulating state. The charge gap can be evaluated by the integrated dip area [33]. The value at $\nu = 1$ (squares) is shown as a function of $E$ (bottom axis) in Fig. 3c.

It is not surprising that the capacitance map resembles that of $R_{xx}$ (Fig. 1c for device 1), minus the signature of the QAH state since compressibility is not affected by edge transport. We identify the QAH state in the capacitance measurements by relying on its characteristic dispersion in doping density and magnetic field (Extended Data Fig. 6). The electric-field span for the QAH state (~ 10 mV/nm) is similar in both devices. Their charge gap from two different measurements is also largely consistent after aligning the critical field for the topological phase transition or the metallic state. As the system approaches the metallic state, the charge gap decreases monotonically. Within our experimental resolution, we do not observe any bulk charge gap closure across the Mott-to-QAH insulator transition.

Next we discuss the electronic band structure and a candidate low-energy model to understand the complex electronic phases in AB-stacked MoTe2/WSe2 heterobilayers. We characterize the moiré bands and the associated wave functions by density functional theory calculations (Methods, Extended Data Fig. 7 and 8). When the bands are non-topological, the wave functions of the first and second dispersive moiré valence bands are centered at the MM and XX sites of the moiré unit cell, respectively, forming a honeycomb lattice structure (Fig. 1a). The out-of-plane electric field tunes the potential difference between the two sites. The moiré band gap at the $K_m/K_m$'-point of the moiré Brillouin zone closes and reopens as $E$ increases (Fig. 1b). A non-zero valley-contrasting Chern number is developed for the bands after the gap reopens. The low-energy physics of the first two moiré bands therefore resembles the Kane-Mele model in the presence of staggered sublattice potential [13]. This is similar to the situation in twisted TMD homobilayers with $E \approx 0$ (Ref. [21,22]). In this model, the $\nu = 2$ state after gap reopening is a quantum valley-spin Hall insulator, which hosts helical edge states. At $\nu = 1$, strong electron correlation can spontaneously stabilize a valley-spin-polarized QAH insulator [21, 22, 34].

The picture is further supported by the behavior of the $\nu = 2$ state in our experiment. The electric-field dependence of $R_{xx}$ at varying temperatures (Fig. 4a) shows a charge gap minimum; the state is insulating on both sides of the minimum. This is consistent with the charge gap from the compressibility measurements at 300 mK (Fig. 4b). The single-particle moiré band gap closes and reopens as $E$ increases. The small non-zero gap at the critical point is presumably from disorder broadening of the transition. We also observe large magnetoresistance immediately after gap closure under an in-plane magnetic field (Fig. 4c, d), but weak magnetoresistance ($\sim 10\%$ at 0.5 T) under an out-of-plane magnetic field with small variations on both sides of the critical point. The behavior is compatible with that of a quantum valley-spin Hall insulator after gap closure [13]. The counter-propagating helical edge states are expected to possess Ising-like spins as in monolayer TMDs; thus only an in-plane magnetic field can mix them to enhance backscattering and induce large magnetoresistance [35]. The presence of helical edge state transport after gap closure is also supported by the non-local transport study in Extended Data Fig. 9. The resistance in our devices is, however, substantially larger than $\frac{h}{2e^2}$ because of backscattering in the helical edge states and bulk dissipation due to the small charge gap.

Finally, we note that the observed absence of bulk gap closure in the Mott-to-QAH insulator transition at $\nu = 1$ is very unusual. The commonly known topological phase transitions involve closing and reopening of a charge gap [13, 23], such as the continuous transition at $\nu = 2$. A topological phase transition without charge gap closure is, however, allowed if the two phases involved in the transition have different symmetries [22, 23, 36, 37]. The QAH insulator here is likely valley-spin-polarized [21, 22, 34]. The Mott insulator could be non-magnetic [38] or has 120-degree Néel order [1]. In the former scenario, time-reversal symmetry is broken only in the QAH insulator. In the latter, the valley pseudospin evolves from a non-collinear to a collinear state. In principle, both scenarios allow a topological phase transition without charge gap closure. Because the QAH insulator has chiral edge states, the transition is likely a weak first-order transition with a small gap discontinuity, which could be smeared out by disorder broadening in our devices.

In conclusion, we have observed a topological phase transition from a Mott insulator to a QAH insulator in AB-stacked MoTe₂/WSe₂ heterobilayers at $\nu = 1$. The transition does not exhibit any sign of charge gap closure and is likely a weak first-order transition. It is correlated with a continuous topological phase transition from a moiré band insulator to a possible quantum valley-spin Hall insulator at $\nu = 2$. Future studies are required to verify the nature of the $\nu = 2$ state after gap closure, as well as the valley-spin-polarized nature of the observed QAH insulator. Our results establish semiconductor moiré materials as a versatile system for exploring the rich phenomenology involving electronic correlation and topology.

**Methods**

**Device fabrications.** We fabricated 60-degree-aligned MoTe$_2$/WSe$_2$ Hall bar devices (Extended Data Fig. 1) using the layer-by-layer dry transfer method as detailed in Ref. [11, 39]. In short, we exfoliated the constituent atomically thin flakes from bulk crystals (HQ Graphene) and picked them up sequentially by a polymer stamp to form the desired stack. The crystal orientations of WSe$_2$ and MoTe$_2$ monolayers and the twist angle between them were determined by the angle-resolved optical second-harmonic generation (SHG) spectroscopy [2, 3]. The angle alignment accuracy is typically ± 0.5°. We used 5-nm-thick Pt as metal electrodes to achieve good electrical contacts to the sample while keeping strain effects minimal. We also used relatively thin (~ 5 nm) hexagonal boron nitride (hBN) gate dielectric in the top gate to achieve a large breakdown electric field of ~ 1 V/nm. We fabricated the capacitance devices following the design reported in Ref. [33]. A global back gate that covers the entire heterobilayer was employed to create a heavily hole-doped contact region. A local top gate was used to deplete the hole density in the region of interest and measure the differential capacitance. We have studied a total of five Hall bar devices. Three devices show the quantized anomalous Hall effect near zero magnetic field at $\nu = 1$. The results for device 1 are presented in the main text. The results of device 2 and 3 are shown in Extended Data Fig. 2.

**Electrical transport measurements.** The electrical transport measurements were performed in a closed-cycle $^4$He cryostat (Oxford TeslatronPT) equipped with a superconducting magnet and a $^3$He insert (base temperature about 300 mK). Low-frequency (< 23 Hz) lock-in techniques were used to measure the sample resistance under a small bias voltage of 1-2 mV to avoid sample heating. The bias current was kept below 15 nA while probing the QAH insulator. Both the voltage drop at the probe electrode pairs and the source-drain current were recorded. Voltage pre-amplifiers with large input impedance (100 MΩ) were used to measure sample resistance up to about 10 MΩ.

**Capacitance measurements.** The capacitance measurements were performed in the same cryostat as the electrical transport measurements. Details have been reported in a recent study of MoSe$_2$/WS$_2$ heterobilayers [33]. In comparison, MoTe$_2$/WSe$_2$ heterobilayers have substantially lower electrical contact resistance; the differential capacitance can be measured at lower temperatures and over a larger frequency range. We used a commercial high electron mobility transistor (HEMT, model FHX35X), which is connected to the sample on the same chip, as a first-stage amplifier [40, 41]; it effectively decouples the device capacitance from the parasitic capacitance from cabling. We measured the differential top-gate capacitance, $C$, by applying an AC voltage (10 mV in amplitude and 3-5 kHz in frequency) to the heterobilayer and collecting charges from the top gate through the HEMT by lock-in techniques. The top-gate capacitance $C$ is related to the quantum capacitance, $C_Q$, of the heterobilayer by $\frac{1}{C} = \frac{1}{C_Q}\left(1 + \frac{C_b}{C_t}\right) + \frac{1}{C_t} \approx \frac{2}{C_Q} + \frac{1}{C_g}$. Here $C_t \approx C_b$ ( $\equiv C_g$) are the geometrical top-gate and back-gate capacitances, respectively, since the two gates are nearly symmetric in the capacitance device (device 6). The chemical potential jump, $\Delta\mu$, at an insulating state can be obtained as $\Delta\mu = e \int dV_t \frac{C_t - C}{C_t + C_b} \approx \frac{e}{2} \int dV_t \left(1 - \frac{C}{C_g}\right)$, where the integration with respect to the top-gate voltage $V_t$ spans the range of the capacitance dip corresponding to the insulating state.

**Band structure calculations and discussions.** Density functional calculations were performed using a generalized gradient approximation [42] with the SCAN+rVV10 van der Waals density functional [43], as implemented in the Vienna Ab initio Simulation Package [44]. Pseudopotentials were used to describe the electron-ion interactions. The structure of 60-degree-twisted (or AB-stacked) $MoTe_2/WSe_2$ heterobilayers was constructed using a vacuum spacing larger than 20 Å to avoid artificial interactions between the periodic images in the out-of-plane direction. Structural relaxation was performed on each atom with force less than 0.01 eV/Å. We used $\Gamma$-point sampling for structural relaxation and self-consistent calculations. The relaxed structure is shown in Extended Data Fig. 7. There are three high-symmetry stacking regions in each moiré unit cell: MM, MX and XX (M = Mo/W, X = Se/Te).

In each TMD monolayer, the band edges are located at the K/K' point of the Brillouin zone. The heterobilayer has a type-I band alignment; both the conduction and valence band edges are from $MoTe_2$ (Ref. [11]). Under zero displacement field, the valence band offset at the K/K' point between the two layers (without the moiré effect) is about 220 meV in the calculations; it is smaller than the experimental value of about 300 meV [11]. The direction of the displacement field $D$ is chosen such that $D$ reduces the valence band offset between the two layers. Our result illustrates the effect of the displacement field on the lowest-energy moiré valence bands. But the absolute values of $D$ cannot be directly compared to that of the applied electric field in the experiment.

Extended Data Fig. 8 shows the moiré valence band dispersions under four representative out-of-plane displacement fields. Under zero displacement field, the top moiré valence band (red) is well separated from the rest moiré bands with a band gap around 10 meV at the $K_m$-point of the moiré Brillouin zone (MBZ). When the displacement field strength is increased, the single-particle band gap between the first two dispersive moiré bands at the $K_m$-point of the MBZ gradually closes around a critical displacement field of 0.25 V/nm; it reopens at higher fields. In addition to the dispersive moiré bands, the figures also show a flat band (blue) that intersects the low-energy dispersive moiré bands.

We analyze the Kohn-Sham wave functions of the moiré bands in real space to identify their origin (Extended Data Fig. 7, 8). We choose two high-symmetry points, the $\Gamma_m$-point and the $K_m$-point of the MBZ. Before the gap closure (e.g. $D = 0.2$ V/nm), the wave functions of each band at the $\Gamma_m$-point and the $K_m$-point are located at the same sites of the moiré unit cell. In particular, the envelope wave function of the flat band is located at the MX site. The atomic-scale wave function consists of both the $d$-orbitals of the M-atoms and $p$-orbitals of the X-atoms; it forms a honeycomb network. This indicates that the flat band is constructed from the $\Gamma$-states of the monolayer TMDs [45]. The envelope wave functions of the first and second dispersive moiré bands are located at the MM and XX sites, respectively. The atomic-scale wave functions are dominated by the $d$-orbitals of the M-atoms and form a triangular network; this indicates that the bands are developed from the K/K' states of the monolayers [45]. The two topmost dispersive moiré bands thus correspond to an effective honeycomb superlattice with MM and XX sublattices.

After gap reopening, the $\Gamma_m$-point wave functions of these bands remain at the same site of the moiré unit cell. However, the $K_m$-point wave functions of the two dispersive bands switch sites, that is, the XX sites now have a higher energy than the MM sites. Our result demonstrates the band inversion and emergence of topological bands with finite valley-resolved Chern numbers after gap reopening. Because the envelope wave functions of the first two dispersive bands form a honeycomb lattice structure, the physics resembles the Kane-Mele model [13, 23]. The displacement field tunes the potential difference between the MM and XX sites and can induce topological moiré bands. This is similar to the case of twisted TMD homobilayers, for which the Kane-Mele-Hubbard model has been proposed [21, 22].

In contrast to the 0-degree-twisted (AA-stacked) MoTe$_2$/WSe$_2$ heterobilayers [11], the interlayer tunneling in AB-stacked heterobilayers is spin-forbidden for the K/K' states in the leading order approximation. The moiré potential is weaker and the moiré bands are more dispersive, as illustrated in Extended Data Fig. 8. The interlayer tunneling is much stronger for the $\Gamma$ states; the moiré potential is stronger and gives rise to the flat band. At charge neutrality around the critical displacement field, the flat band slightly overlaps with the first dispersive moiré band. However, with finite hole doping between $\nu = 1$ and $\nu = 2$, the flat band will be pushed away by electrostatic repulsion, and therefore can be ignored in the transport measurements.

## References

1.   Wu, F., Lovorn, T., Tutuc, E. & MacDonald, A.H. Hubbard Model Physics in Transition Metal Dichalcogenide Moir\'e Bands. *Physical Review Letters* **121**, 026402 (2018).

2.   Regan, E.C., Wang, D., Jin, C., Bakti Utama, M.I., Gao, B., Wei, X., Zhao, S., Zhao, W., Zhang, Z., Yumigeta, K., Blei, M., Carlström, J.D., Watanabe, K., Taniguchi, T., Tongay, S., Crommie, M., Zettl, A. & Wang, F. Mott and generalized Wigner crystal states in WSe2/WS2 moiré superlattices. *Nature* **579**, 359-363 (2020).

3.   Tang, Y., Li, L., Li, T., Xu, Y., Liu, S., Barmak, K., Watanabe, K., Taniguchi, T., MacDonald, A.H. & Shan, J. Simulation of Hubbard model physics in WSe 2/WS 2 moiré superlattices. *Nature* **579**, 353-358 (2020).

4.   Wang, L., Shih, E.-M., Ghiotto, A., Xian, L., Rhodes, D.A., Tan, C., Claassen, M., Kennes, D.M., Bai, Y., Kim, B., Watanabe, K., Taniguchi, T., Zhu, X., Hone, J., Rubio, A., Pasupathy, A.N. & Dean, C.R. Correlated electronic phases in twisted bilayer transition metal dichalcogenides. *Nature Materials* **19**, 861-866 (2020).

5.   Shimazaki, Y., Schwartz, I., Watanabe, K., Taniguchi, T., Kroner, M. & Imamoğlu, A. Strongly correlated electrons and hybrid excitons in a moiré heterostructure. *Nature* **580**, 472-477 (2020).

6.   Xu, Y., Liu, S., Rhodes, D.A., Watanabe, K., Taniguchi, T., Hone, J., Elser, V., Mak, K.F. & Shan, J. Correlated insulating states at fractional fillings of moiré superlattices. *Nature* **587**, 214-218 (2020).

7. Zhang, Y., Yuan, N.F.Q. & Fu, L. Moir\'e quantum chemistry: Charge transfer in transition metal dichalcogenide superlattices. *Physical Review B* **102**, 201115 (2020).

8. Andrei, E.Y., Efetov, D.K., Jarillo-Herrero, P., MacDonald, A.H., Mak, K.F., Senthil, T., Tutuc, E., Yazdani, A. & Young, A.F. The marvels of moiré materials. *Nature Reviews Materials* **6**, 201-206 (2021).

9. Huang, X., Wang, T., Miao, S., Wang, C., Li, Z., Lian, Z., Taniguchi, T., Watanabe, K., Okamoto, S., Xiao, D., Shi, S.-F. & Cui, Y.-T. Correlated insulating states at fractional fillings of the WS2/WSe2 moiré lattice. *Nature Physics* **17**, 715-719 (2021).

10. Jin, C., Tao, Z., Li, T., Xu, Y., Tang, Y., Zhu, J., Liu, S., Watanabe, K., Taniguchi, T., Hone, J.C., Fu, L., Shan, J. & Mak, K.F. Stripe phases in WSe2/WS2 moiré superlattices. *Nature Materials* **20**, 940-944 (2021).

11. Li, T., Jiang, S., Li, L., Zhang, Y., Kang, K., Zhu, J., Watanabe, K., Taniguchi, T., Chowdhury, D. & Fu, L. Continuous Mott transition in semiconductor moir\'e superlattices. *arXiv preprint arXiv:2103.09779* (2021).

12. Augusto Ghiotto, En-Min Shih, Giancarlo S. S. G. Pereira, Daniel A. Rhodes, Bumho Kim, Jiawei Zang, Andrew J. Millis, Kenji Watanabe, Takashi Taniguchi, James C. Hone, Lei Wang, Cory R. Dean & Pasupathy, A.N. Quantum Criticality in Twisted Transition Metal Dichalcogenides. *arXiv:2103.09796* (2021).

13. Hasan, M.Z. & Kane, C.L. Colloquium: Topological insulators. *Reviews of Modern Physics* **82**, 3045-3067 (2010).

14. Andrei, E.Y. & MacDonald, A.H. Graphene bilayers with a twist. *Nature Materials* **19**, 1265-1275 (2020).

15. Balents, L., Dean, C.R., Efetov, D.K. & Young, A.F. Superconductivity and strong correlations in moiré flat bands. *Nature Physics* **16**, 725-733 (2020).

16. Kennes, D.M., Claassen, M., Xian, L., Georges, A., Millis, A.J., Hone, J., Dean, C.R., Basov, D.N., Pasupathy, A.N. & Rubio, A. Moiré heterostructures as a condensed-matter quantum simulator. *Nature Physics* **17**, 155-163 (2021).

17. Serlin, M., Tschirhart, C.L., Polshyn, H., Zhang, Y., Zhu, J., Watanabe, K., Taniguchi, T., Balents, L. & Young, A.F. Intrinsic quantized anomalous Hall effect in a moiré heterostructure. *Science* **367**, 900 (2020).

18. Sharpe, A.L., Fox, E.J., Barnard, A.W., Finney, J., Watanabe, K., Taniguchi, T., Kastner, M.A. & Goldhaber-Gordon, D. Emergent ferromagnetism near three-quarters filling in twisted bilayer graphene. *Science* **365**, 605 (2019).

19. Chen, G., Sharpe, A.L., Fox, E.J., Zhang, Y.-H., Wang, S., Jiang, L., Lyu, B., Li, H., Watanabe, K., Taniguchi, T., Shi, Z., Senthil, T., Goldhaber-Gordon, D., Zhang, Y. & Wang, F. Tunable correlated Chern insulator and ferromagnetism in a moiré superlattice. *Nature* **579**, 56-61 (2020).

20. Cao, Y., Fatemi, V., Fang, S., Watanabe, K., Taniguchi, T., Kaxiras, E. & Jarillo-Herrero, P. Unconventional superconductivity in magic-angle graphene superlattices. *Nature* **556**, 43-50 (2018).

21. Wu, F., Lovorn, T., Tutuc, E., Martin, I. & MacDonald, A.H. Topological Insulators in Twisted Transition Metal Dichalcogenide Homobilayers. *Physical Review Letters* **122**, 086402 (2019).

22. Trithep Devakul, Valentin Crépel, Yang Zhang & Fu, L. Magic in twisted transition metal dichalcogenide bilayers. *arXiv:2106.11954* (2021).

23. Hohenadler, M. & Assaad, F.F. Correlation effects in two-dimensional topological insulators. *Journal of Physics: Condensed Matter* **25**, 143201 (2013).

24. Witczak-Krempa, W., Chen, G., Kim, Y.B. & Balents, L. Correlated Quantum Phenomena in the Strong Spin-Orbit Regime. *Annual Review of Condensed Matter Physics* **5**, 57-82 (2014).

25. Pesin, D. & Balents, L. Mott physics and band topology in materials with strong spin–orbit interaction. *Nature Physics* **6**, 376-381 (2010).

26. Raghu, S., Qi, X.-L., Honerkamp, C. & Zhang, S.-C. Topological Mott Insulators. *Physical Review Letters* **100**, 156401 (2008).

27. Chang, C.-Z., Zhang, J., Feng, X., Shen, J., Zhang, Z., Guo, M., Li, K., Ou, Y., Wei, P., Wang, L.-L., Ji, Z.-Q., Feng, Y., Ji, S., Chen, X., Jia, J., Dai, X., Fang, Z., Zhang, S.-C., He, K., Wang, Y., Lu, L., Ma, X.-C. & Xue, Q.-K. Experimental Observation of the Quantum Anomalous Hall Effect in a Magnetic Topological Insulator. *Science* **340**, 167 (2013).

28. Kou, X., Guo, S.-T., Fan, Y., Pan, L., Lang, M., Jiang, Y., Shao, Q., Nie, T., Murata, K., Tang, J., Wang, Y., He, L., Lee, T.-K., Lee, W.-L. & Wang, K.L. Scale-Invariant Quantum Anomalous Hall Effect in Magnetic Topological Insulators beyond the Two-Dimensional Limit. *Physical Review Letters* **113**, 137201 (2014).

29. Mogi, M., Yoshimi, R., Tsukazaki, A., Yasuda, K., Kozuka, Y., Takahashi, K.S., Kawasaki, M. & Tokura, Y. Magnetic modulation doping in topological insulators toward higher-temperature quantum anomalous Hall effect. *Applied Physics Letters* **107**, 182401 (2015).

30. Liu, C.-X., Zhang, S.-C. & Qi, X.-L. The Quantum Anomalous Hall Effect: Theory and Experiment. *Annual Review of Condensed Matter Physics* **7**, 301-321 (2016).

31. MacDonald, A.H. Introduction to the Physics of the Quantum Hall Regime. *arXiv:cond-mat/9410047* (1994).

32. Young, A.F., Sanchez-Yamagishi, J.D., Hunt, B., Choi, S.H., Watanabe, K., Taniguchi, T., Ashoori, R.C. & Jarillo-Herrero, P. Tunable symmetry breaking and helical edge transport in a graphene quantum spin Hall state. *Nature* **505**, 528-532 (2014).

33. Li, T., Zhu, J., Tang, Y., Watanabe, K., Taniguchi, T., Elser, V., Shan, J. & Mak, K.F. Charge-order-enhanced capacitance in semiconductor moir\'e superlattices. *arXiv preprint arXiv:2102.10823* (2021).

34. Ying-Ming Xie, Cheng-Ping Zhang, Jin-Xin Hu, Kin Fai Mak & Law, K.T. Theory of Valley Polarized Quantum Anomalous Hall State in Moiré MoTe2/WSe2 Heterobilayers. *arXiv:2106.13991* (2021).

35. Fei, Z., Palomaki, T., Wu, S., Zhao, W., Cai, X., Sun, B., Nguyen, P., Finney, J., Xu, X. & Cobden, D.H. Edge conduction in monolayer WTe2. *Nature Physics* **13**, 677-682 (2017).

36. Amaricci, A., Budich, J.C., Capone, M., Trauzettel, B. & Sangiovanni, G. First-Order Character and Observable Signatures of Topological Quantum Phase Transitions. *Physical Review Letters* **114**, 185701 (2015).

37.     Ezawa, M., Tanaka, Y. & Nagaosa, N. Topological Phase Transition without Gap Closing. *Scientific Reports* **3**, 2790 (2013).

38.     Senthil, T. Theory of a continuous Mott transition in two dimensions. *Physical Review B* **78**, 045109 (2008).

39.     Wang, L., Meric, I., Huang, P.Y., Gao, Q., Gao, Y., Tran, H., Taniguchi, T., Watanabe, K., Campos, L.M., Muller, D.A., Guo, J., Kim, P., Hone, J., Shepard, K.L. & Dean, C.R. One-Dimensional Electrical Contact to a Two-Dimensional Material. *Science* **342**, 614 (2013).

40.     Zibrov, A.A., Kometter, C., Zhou, H., Spanton, E.M., Taniguchi, T., Watanabe, K., Zaletel, M.P. & Young, A.F. Tunable interacting composite fermion phases in a half-filled bilayer-graphene Landau level. *Nature* **549**, 360-364 (2017).

41.     Ashoori, R.C., Stormer, H.L., Weiner, J.S., Pfeiffer, L.N., Pearton, S.J., Baldwin, K.W. & West, K.W. Single-electron capacitance spectroscopy of discrete quantum levels. *Physical Review Letters* **68**, 3088-3091 (1992).

42.     Perdew, J.P., Burke, K. & Ernzerhof, M. Generalized Gradient Approximation Made Simple. *Physical Review Letters* **77**, 3865-3868 (1996).

43.     Peng, H., Yang, Z.-H., Perdew, J.P. & Sun, J. Versatile van der Waals Density Functional Based on a Meta-Generalized Gradient Approximation. *Physical Review X* **6**, 041005 (2016).

44.     Kresse, G. & Furthmüller, J. Efficiency of ab-initio total energy calculations for metals and semiconductors using a plane-wave basis set. *Computational Materials Science* **6**, 15-50 (1996).

45.     Liu, G.-B., Xiao, D., Yao, Y., Xu, X. & Yao, W. Electronic structures and theoretical modelling of two-dimensional group-VIB transition metal dichalcogenides. *Chemical Society Reviews* **44**, 2643-2663 (2015).

**Figures**

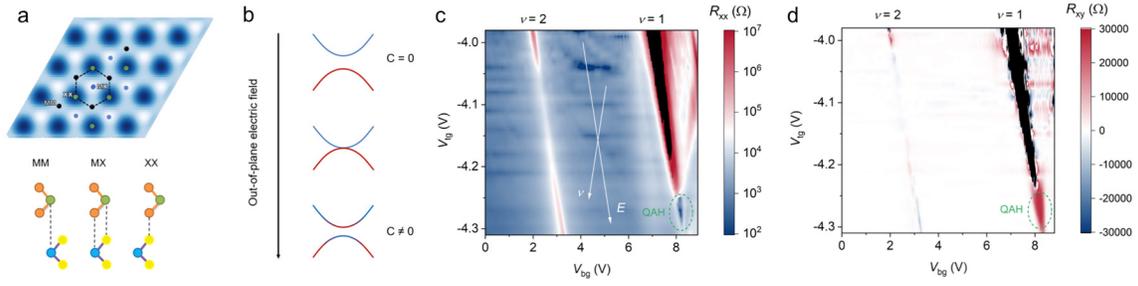

**Figure 1 | AB-stacked MoTe₂/WSe₂ heterobilayer. a,** Top: triangular moiré superlattice with high-symmetry stacking sites MM, MX and XX (M = Mo/W; X = Se/Te). The interlayer distance increases from the white to the blue region. MM and XX form a honeycomb structure (dashed line). Bottom: orange and yellow balls denote X atoms; green and blue balls denote M atoms; dashed lines show vertical atomic alignment. **b,** Schematic illustration of controlling band topology by an out-of-plane electric field. As the field increases, the gap between two dispersive moiré bands closes and reopens. Topologically nontrivial bands ($c \neq 0$) emerge after gap reopening. **c,d,** Longitudinal (**c**) and Hall (**d**) resistance at 300 mK as a function of top and bottom gate voltages. $R_{xx}$ is measured under zero magnetic field; $R_{xy}$ is the symmetrized result under an out-of-plane magnetic field of ±0.1 T. The two gate voltages vary the filling factor ($\nu$) and electric field ($E$) independently along the white arrow directions. Dashed circle mark the region of a QAH insulator.

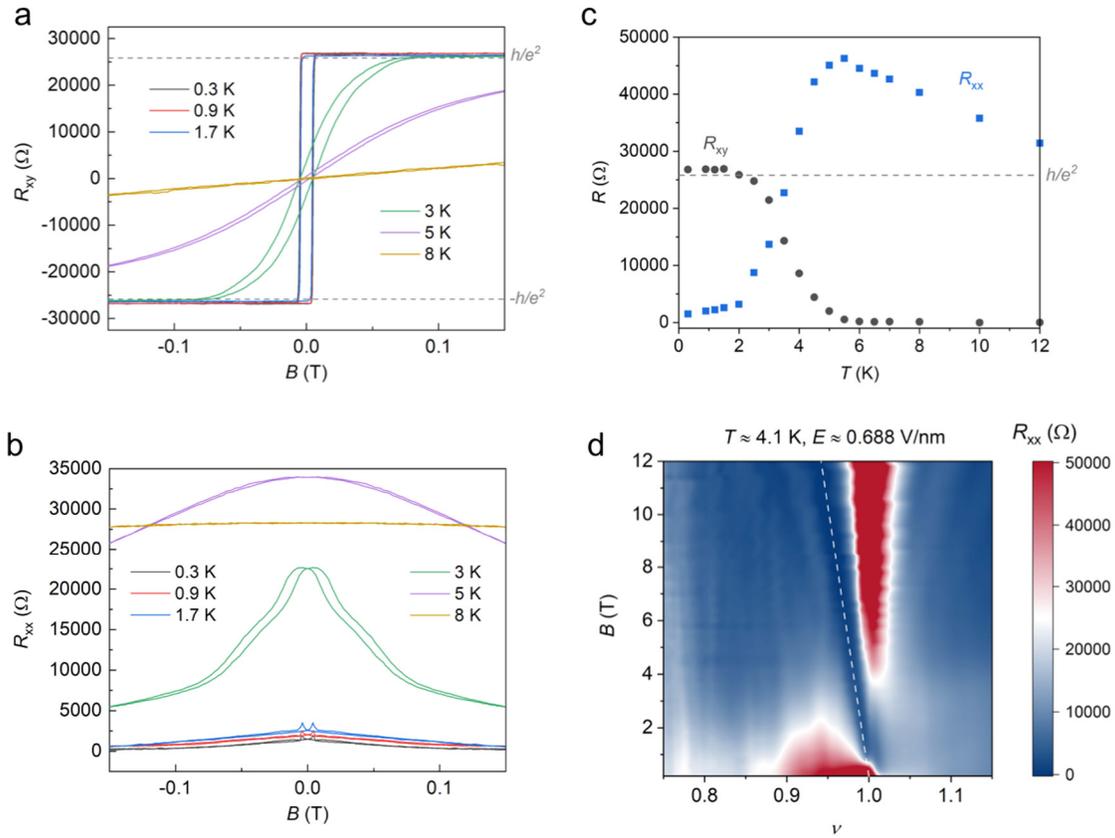

**Figure 2 | Quantized anomalous Hall effect at $\nu = 1$. a,b,** Magnetic-field dependence of $R_{xy}$ (**a**) and $R_{xx}$ (**b**) of the QAH insulator at varying temperatures. Quantized Hall resistance ($R_{xy} \approx \frac{h}{e^2}$) and small longitudinal resistance ($R_{xx} \ll R_{xy}$), accompanied by a negative magnetoresistance, are observed at the three lowest temperatures. There is also a clear magnetic hysteresis with a sharp magnetic switching near ± 10 mT. **c,** Temperature dependence of the zero-magnetic-field value of $R_{xy}$ and $R_{xx}$. The onset of magnetic ordering is at ~ 5-6 K and quantization of $R_{xy}$ happens below ~ 2.5 K. The dependences of $R_{xy}$ and $R_{xx}$ are perfectly correlated. **d,** $R_{xx}$ as a function of magnetic field and filling factor at 4.1 K. The location of the $R_{xx}$ minimum (dashed line) disperses with magnetic field with a slope corresponds to a Chern number 1.

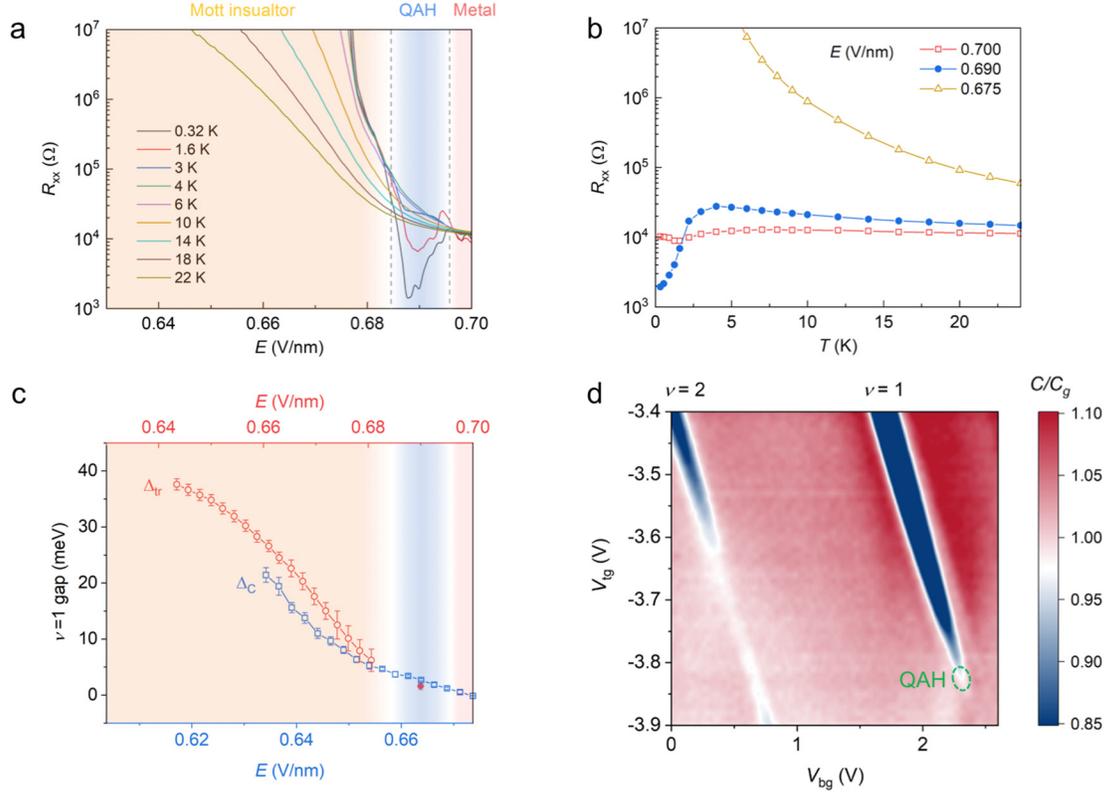

**Figure 3 | Mott-to-QAH insulator transition at ν = 1. a,** Electric-field dependence of $R_{xx}$ at ν = 1 under zero magnetic field at varying temperatures (device 1). Three regions, corresponding to a Mott insulator, QAH insulator and metal, are identified. **b,** Representative temperature dependences of $R_{xx}$, one from each region in **a**. **c,** Charge gap as a function of electric field extracted from thermal activation transport (red circles, device 1, top axis) and extracted from capacitance measurements at 300 mK (blue squares, device 6, bottom axis). The error bars for the red symbols are uncertainties of the activation fit; the error bars for the blue symbols are propagated uncertainties of the capacitance measurement and numerical integration of the capacitance dips. The two electric field axes are slightly shifted to match the phase boundaries in two devices. The charge gap evolves smoothly across the topological phase transition. **d,** Differential top-gate capacitance at 300 mK and at zero magnetic field as a function of top and bottom gate voltages (device 6). The two prominent incompressible states correspond to ν = 1 and ν = 2. The green dashed line denotes the QAH region, identified by the magnetic-field dispersion of the state (Extended Data Fig. 6). The lines in **b**, **c** are guides to the eye.

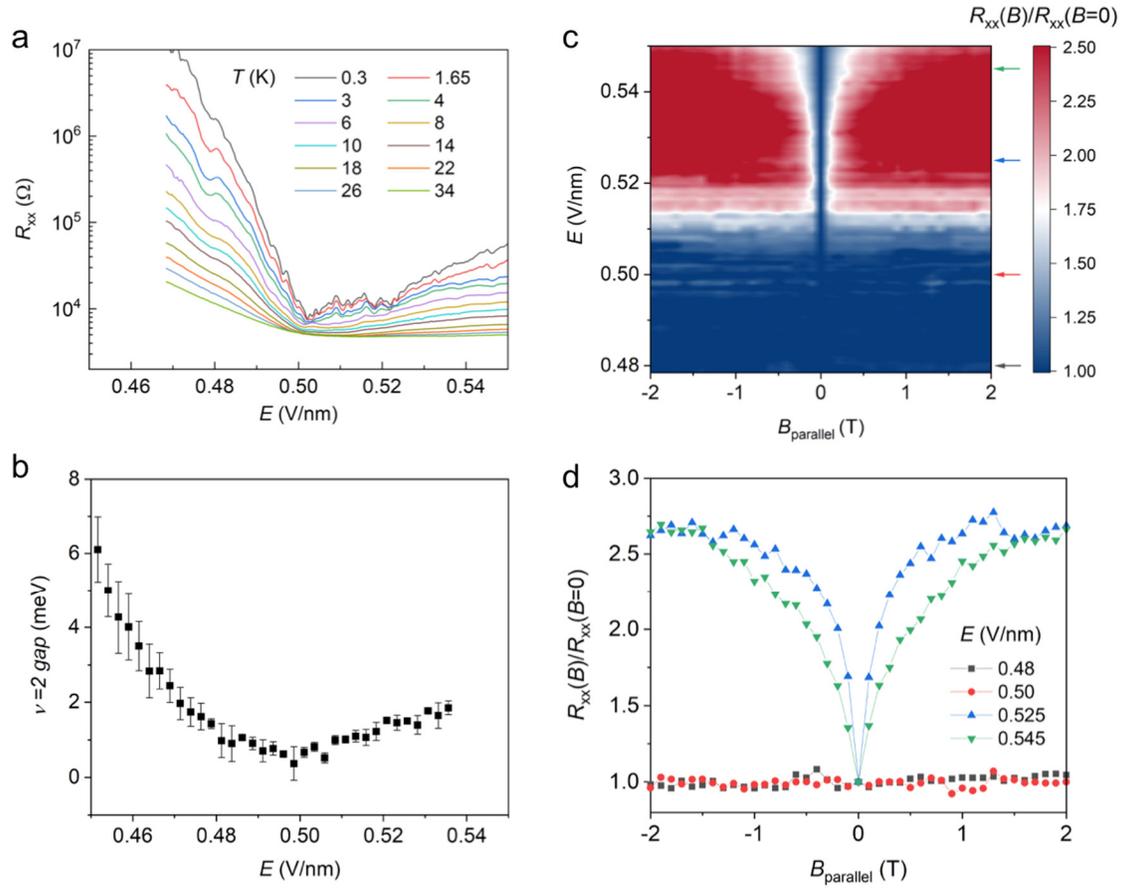

**Figure 4 | Evidence of a quantum valley-spin Hall insulator at $\nu = 2$. a,** Electric-field dependence of $R_{xx}$ at $\nu = 2$ under zero magnetic field for varying temperatures (device 1). The temperature dependence becomes much weaker after gap reopening. **b,** Charge gap at $\nu = 2$ from capacitance measurements at 300 mK (device 6). As the electric field increases, the charge gap closes and reopens. The gap size is much smaller after the gap reopens. The error bars are propagated uncertainties of the capacitance measurement and numerical integration of the capacitance dips. **c,** Electric-field dependence of magnetoresistance, $\frac{R_{xx}(B)}{R_{xx}(B=0)}$, under an in-plane magnetic field at 300 mK. **d,** Magnetoresistance (symbols) at selected electric fields denoted by arrows in **c**. The lines are guides to the eye. Strong magnetoresistance is observed only after the gap reopens.

**Extended data figures**

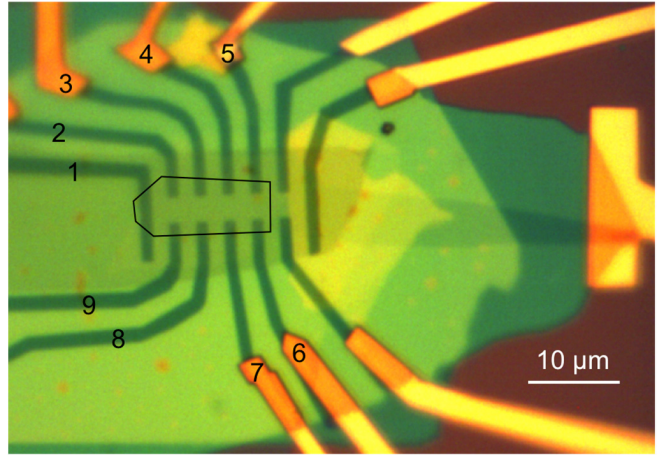

**Extended Data Figure 1 | Optical micrograph of device 1.** MoTe$_2$/WSe$_2$ heterobilayer is outlined by the black line. Electrode 1-9 are labeled. For the results presented in the main text, electrode 5 and 6 are grounded; electrode 1 is used as a source electrode. The longitudinal voltage drop is measured between 3 and 4; and the transverse voltage drop is measured between 3 and 8. The scale bar is 10 μm.

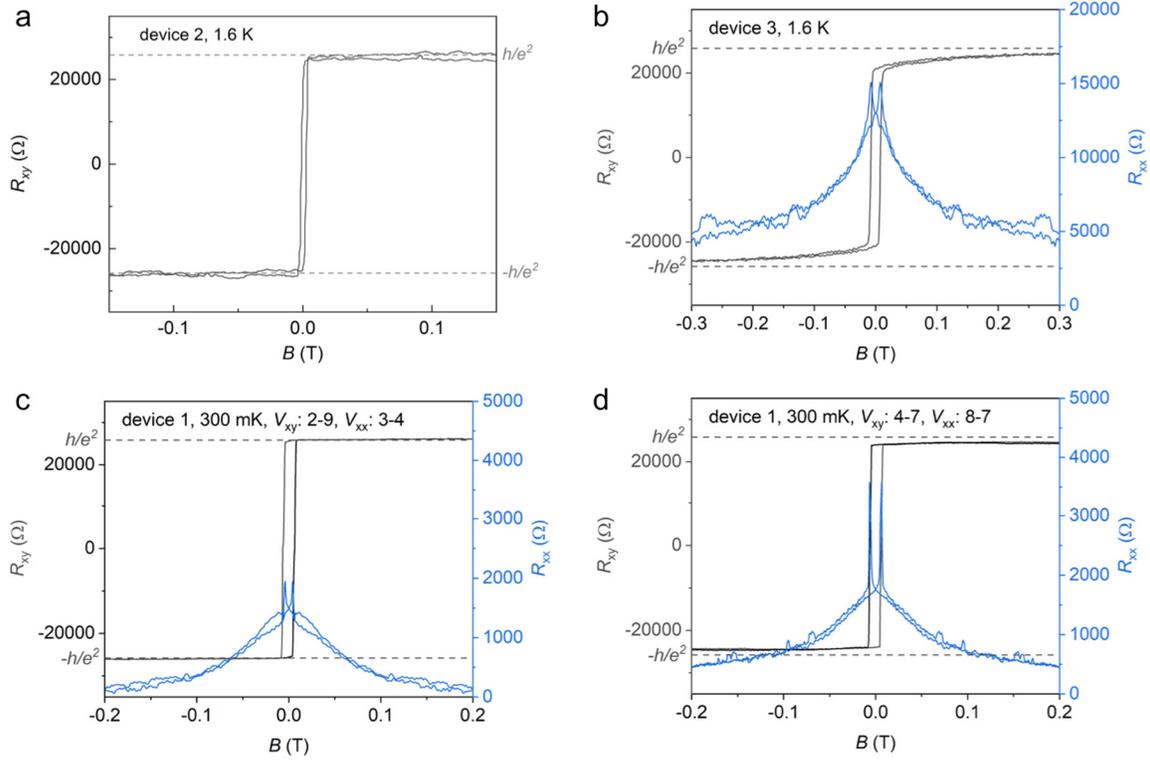

**Extended Data Figure 2 | Other devices exhibiting the QAH effect at $\nu = 1$. a,b,** Magnetic-field dependence of $R_{xy}$ and $R_{xx}$ from device 2 and 3 at 1.6 K. Quantized Hall resistance at zero magnetic field is observed in device 2; $R_{xx}$ is not measured. The resistance quantization in device 3 is nearly perfect at a small magnetic field, accompanied by a significant drop in $R_{xx}$. **c,d,** The same as **a,b** for device 1 measured from different pairs of probes (than the result shown in the main text) at 300 mK. The labeling of the electrodes is shown in Extended Data Fig. 1. Nearly quantized resistance with similar magnetic-field dependence is observed. The coercive field varies slightly from device to device.

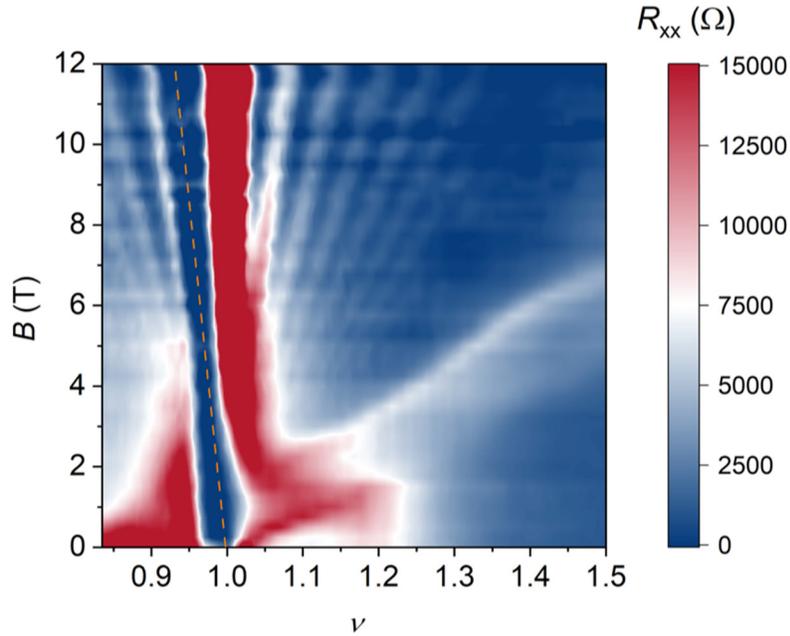

**Extended Data Figure 3 | Landau fan at 300 mK.** Longitudinal resistance as a function of magnetic field and filling factor (device 1). The $R_{xx}$ minimum, denoted by the red dashed line, corresponds to the QAH insulator; it is present at zero magnetic field and corresponds to a Chern number 1. In addition to the QAH insulator, a Landau fan emerges from $\nu = 1$. The Landau level degeneracy is 1, consistent with the valley-spin-polarized Landau levels.

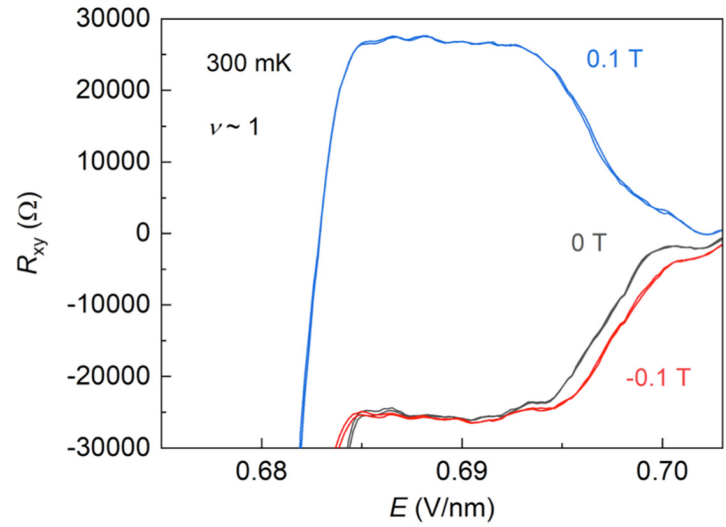

**Extended Data Figure 4 | Absence of electric-field hysteresis.** Hall resistance at $\nu = 1$ as a function of electric field under both forward and backward scans at 300 mK (device 1). The out-of-plane magnetic field is fixed at 0 or ±0.1 T. No hysteresis is observed.

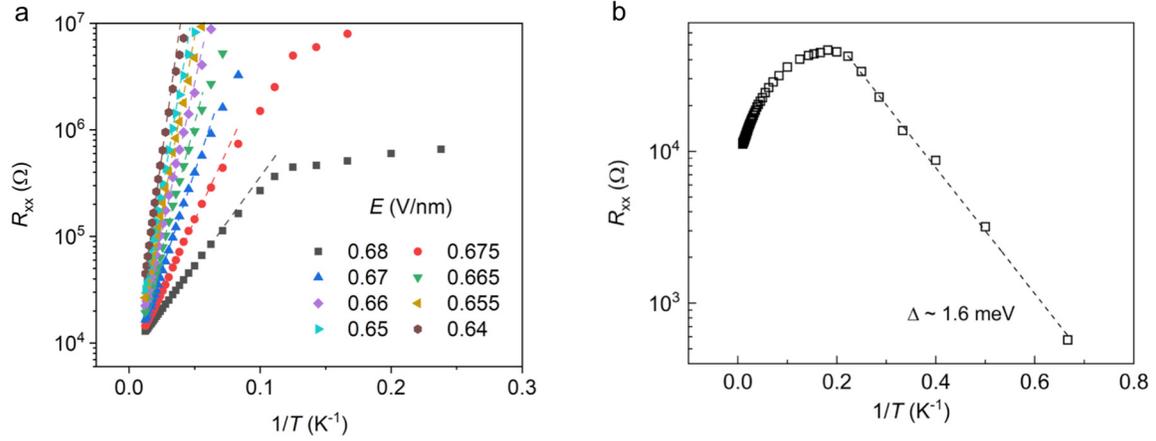

**Extended Data Figure 5 | Charge gap from thermal activation transport. a,** Arrhenius plot of $R_{xx}$ at varying electric fields for the Mott insulator. The high-temperature data is well described by the thermal activation model (dashed lines), from which the activation gap is extracted. **b,** Same as **a** for the QAH insulator. Here thermal activation behavior (dashed line) is observed only at low temperature, when the QAH state develops. The extracted charge gaps are shown as empty red circles (Mott insulator) and filled circle (QAH insulator) in Fig. 3c of the main text.

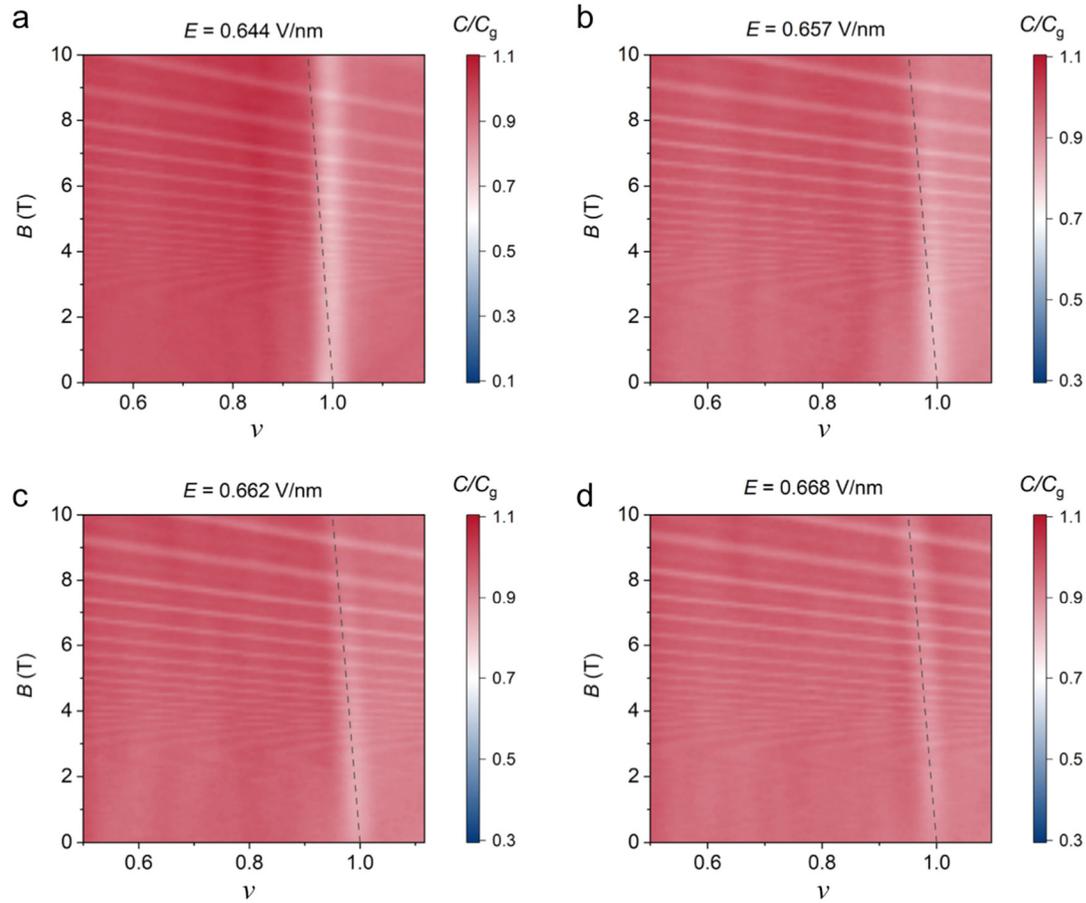

**Extended Data Figure 6 | Identification of the QAH region in the capacitance device (device 6). a-d,** Normalized differential top-gate capacitance, $C/C_g$, versus magnetic field and filling factor at 300 mK. Results for four different electric fields are shown. Most strongly dispersive incompressible states shown arise from the Landau levels of the graphite top gate and are irrelevant in this study. The black dashed lines, originated from $\nu = 1$, are the theoretical magnetic-field dispersion of a QAH state with Chern number 1. We use this to determine the Mott-QAH insulator boundary in Fig. 3c of the main text. The system is a Mott insulator in **a** since there is no magnetic field dispersion of the incompressible state; it is near the Mott-QAH insulator boundary in **b**; and it is a QAH insulator in **c**, **d**. The QAH insulator-metal boundary is determined as capacitance reaches $C/C_g \approx 1$ in the metallic state.

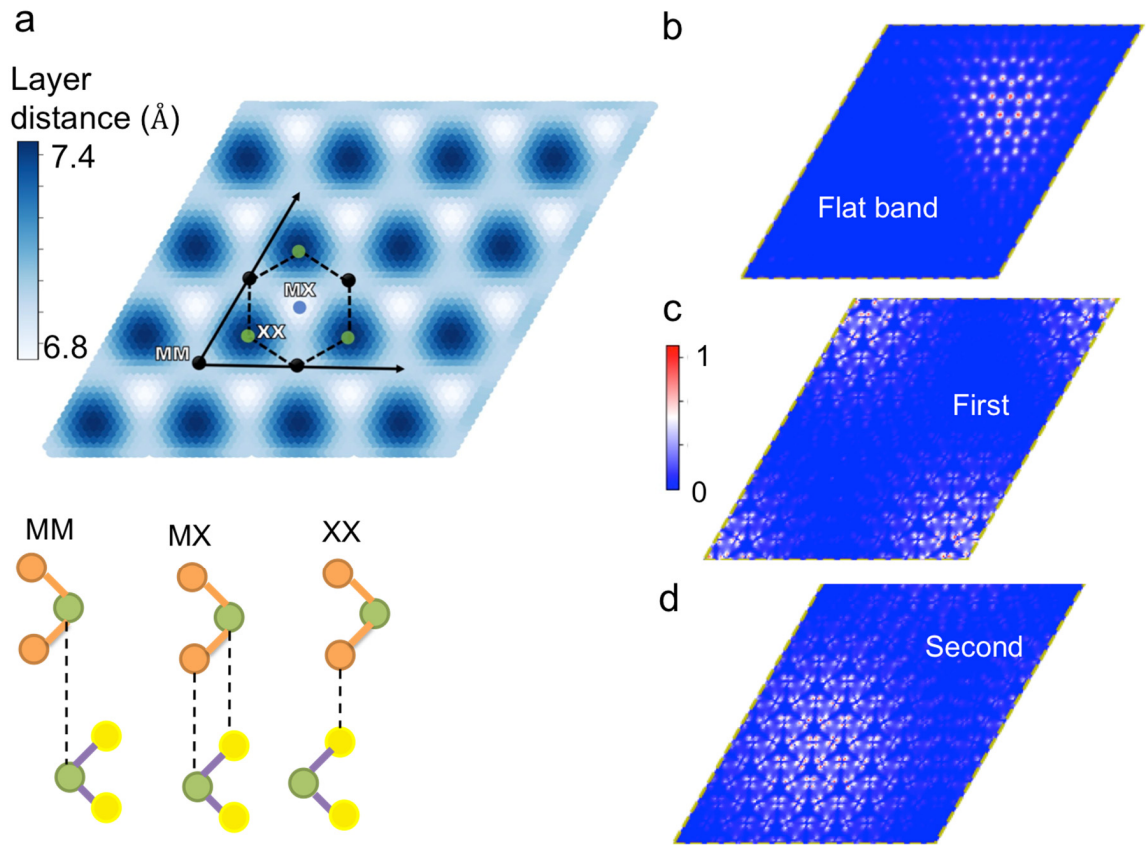

**Extended Data Figure 7 | Kohn-Sham wave functions. a,** Interlayer distance (top) and stacking alignment (bottom) of relaxed AB-stacked $MoTe_2/WSe_2$ heterobilayers. MM, MX, and XX (M = Mo/W, X = Se/Te) are the high-symmetry stacking sites. **b-d,** Real-space wave function at the $\Gamma_m$-point of the MBZ for the flat band (**b**), the first (**c**) and second (**d**) dispersive valence moiré bands. The envelope wave function for the flat band is located at the MX site. The atomic-scale orbitals consist of both the $d$-orbitals of the M atom and the $p$-orbitals of the X atom, forming a honeycomb network. The wave functions for the first and second dispersive moiré bands are located at the MM and XX sites, respectively. The atomic-scale orbitals are dominated by the $d$-orbitals of the M atom, forming a triangular network. The results show that the dispersive bands are originated from the K/K' valleys of the TMD monolayers.

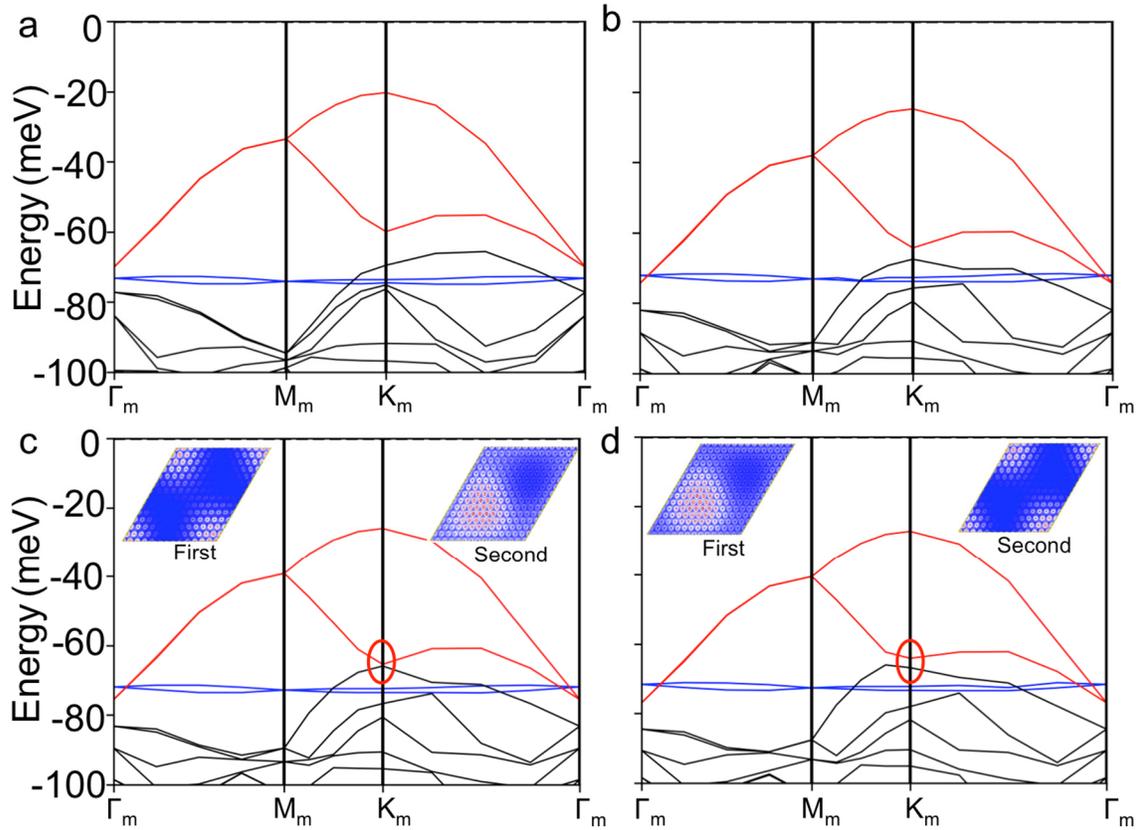

**Extended Data Figure 8 | Moiré band structure at varying displacement fields.**
Moiré valence band structure of AB-stacked MoTe$_2$/WSe$_2$ heterobilayers from DFT with
displacement field 0 V/nm (**a**), 0.2 V/nm (**b**), 0.25 V/nm (**c**) and 0.3 V/nm (**d**). The first
moiré band is shown in red; the flat band is in blue; and the rest of the bands are in black.
The energy gap at the K$_m$-point (marked by a red circle) between the two dispersive moiré bands
closes and reopens between 0.25 V/nm and 0.3 V/nm. The insets in **c** and **d**
show the envelope wave function at the K$_m$-point of the MBZ for the first (left inset) and
second (right inset) dispersive bands. Before gap closure, the K$_m$-point wave function for
the two dispersive bands is centered at the MM and XX site, respectively. They are
centered at the same sites as the Γ$_m$-point wave function (see Extended Data Fig. 7). The
envelope wave function switches moiré sites after gap reopening, showing the emergence
of topological bands. See Methods for the role of the flat band.

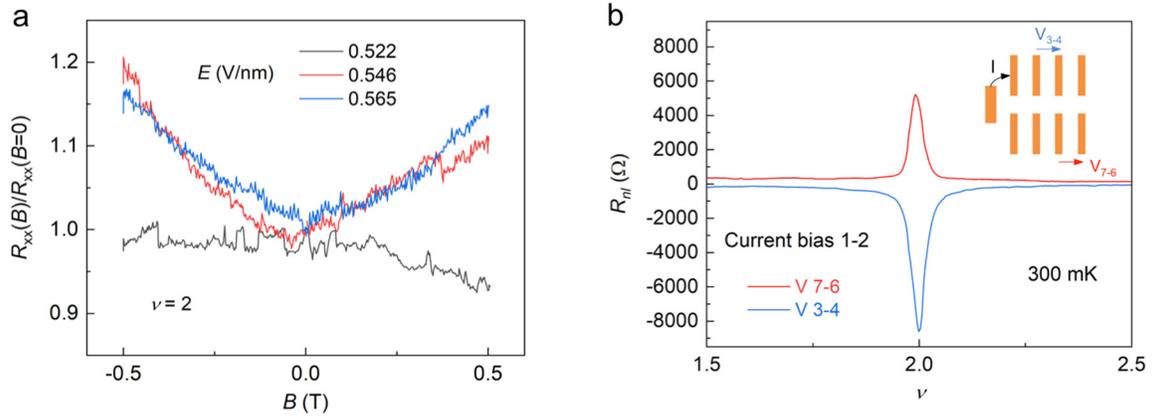

**Extended Data Figure 9 | Magnetoresistance and nonlocal transport at $\nu = 2$. a,** Magnetoresistance, $\frac{R_{xx}(B)}{R_{xx}(B=0)}$, at 300 mK under an out-of-plane magnetic field at varying electric fields. The effect is much weaker than magnetoresistance under an in-plane magnetic field (Fig. 4c, d of the main text). **b,** Nonlocal resistance versus filling factor around $\nu = 2$ after gap reopening. The magnetic field is zero. The arrow for I in the inset shows the direction of the bias electric field between the source and drain electrodes. Voltage drops between electrode 3 and 4 ($V_{3\text{-}4}$) and between 7 and 6 ($V_{7\text{-}6}$) are measured. A change in sign for the nonlocal resistance is consistent with the presence of helical edge state transport for a quantum valley-spin Hall insulator.